# Scattering of Cold Atom Coherences by Hot Atoms: Frequency Shifts from Background Gas Collisions

Kurt Gibble
Department of Physics, The Pennsylvania State University, University Park, PA 16802

Frequency shifts from background gas collisions currently contribute significantly to the inaccuracy of atomic clocks. Because nearly all collisions with room-temperature background gases that transfer momentum eject the cold atoms from the clock, the interference between the scattered and unscattered waves in the forward direction dominates these frequency shifts. We show they are ≈10 times smaller than in room-temperature clocks and that van der Waals interactions produce the cold-atom background-gas shift. General considerations allow the loss of Ramsey fringe amplitude to bound this frequency shift.

PACS: 06.30.Ft, 34.50.Cx

The frequency shift due to background gas collisions is currently a leading systematic uncertainty in the best atomic clocks, which have accuracies of $2.1\times10^{-16}$ [1,2]. Coherent superpositions are central to atomic clocks and atom interferometers for precision measurements [3,4]. While the underlying scattering theory is thoroughly established, the scattering physics of coherences [5] continues to unfold – recent examples include phase shifts of scattered coherences [6] and interactions when non-identical coherent superpositions collide [7-9]. Here we treat the scattering of cold clock atoms, coherent superpositions of two internal states, by a background gas at room temperature.

Current laser-cooled fountain clocks bound the frequency shift from background gas collisions using measurements from room-temperature clocks [10]. This is not justified since any momentum transferred to a cold atom almost always prevents it from being detected (Fig. 1). A first guess is that the frequency shifts of the remaining atoms must be much smaller. If so, does a frequency shift remain? Ion clocks that use quantum-logic to readout the atom's trap state are a particularly striking example [11] since any momentum transfer destroys the detected coherence. We show that the dominant frequency shift for cold atoms has no momentum transfer. The scale of the shift is smaller than the shift used from room-temperature clocks and has the opposite sign. Interestingly, the shift can be directly related to the decrease in Ramsey fringe amplitude due to background gas collisions, with the same dependences, giving another technique to evaluate and reduce this currently significant systematic error. We treat the quantum scattering for cold atom coherences and show that they are only sensitive to the weak, usually long-range, collisions, while room-temperature clocks are sensitive to both short and long-range collisions. Polarizable gases give large contributions, motivating a reduction [2] and evaluation of shifts from background cesium vapor. We also discuss the implications for ion [11] and optical lattice clocks [12-14], which trap cold, coherent superpositions.

When an atom, in a superposition of two internal clock states, collides with a perturber, the scattered outgoing wave contains a coherent superposition of the two clock states, each with an angle-dependent scattering amplitude, and the perturber. Because, the scattering amplitude is generally different for the two clock states, their coherent superposition experiences a phase shift; each clock state can even scatter in classically different directions, leading to a loss of coherence [5].

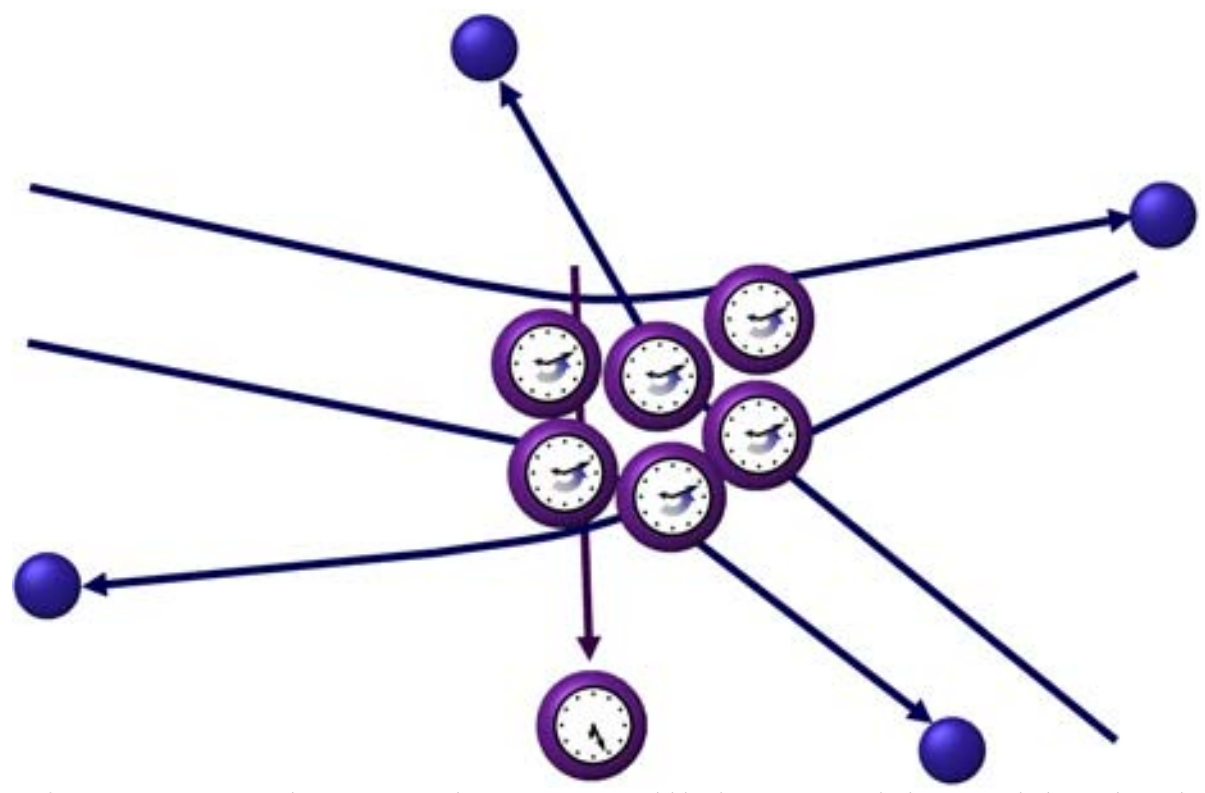

**Fig. 1.** Background gas collisions with cold clock atoms eject clock atoms so that they are not detected. Nonetheless, undeflected cold atom coherences experience a phase shift due to the scattering. The frequency shifts of undeflected cold clock-atoms are due to long-range collisions with polarizable background gases.

General behaviors of the background gas scattering of cold clock atoms emerge from a partial angular-momentum wave expansion. It leads to three contributions to the outgoing probability current, an unscattered wave, a scattered wave, and an interference between the scattered and unscattered waves in exactly the forward direction, $\theta=0$. For cold atoms, which have negligible momentum in their rest frame, we show below that the interference term ($\theta=0$) contributes much more than the scattered wave, even though the scattered wave is highly peaked near $\theta\approx0$ [5,15-17]. For the usual clock Ramsey spectroscopy, two separated $\pi/2$ pulses with a relative phase $\phi$, the interference current of the excited state $|2\rangle$ is $j_{2,int}=\pi\hbar/\mu(\mathrm{Re}[f_1(0)-f_2(0)]\sin(\phi)-2\mathrm{Im}[f_1(0)+f_2(0)]\cos^2(\phi/2))$, as outlined below. Using the scattering amplitudes $f_\gamma(\theta)=\Sigma(2\ell+1)[\exp(2i\delta_{\gamma\ell})-1]P_\ell[\cos(\theta)]/2ik$ gives:

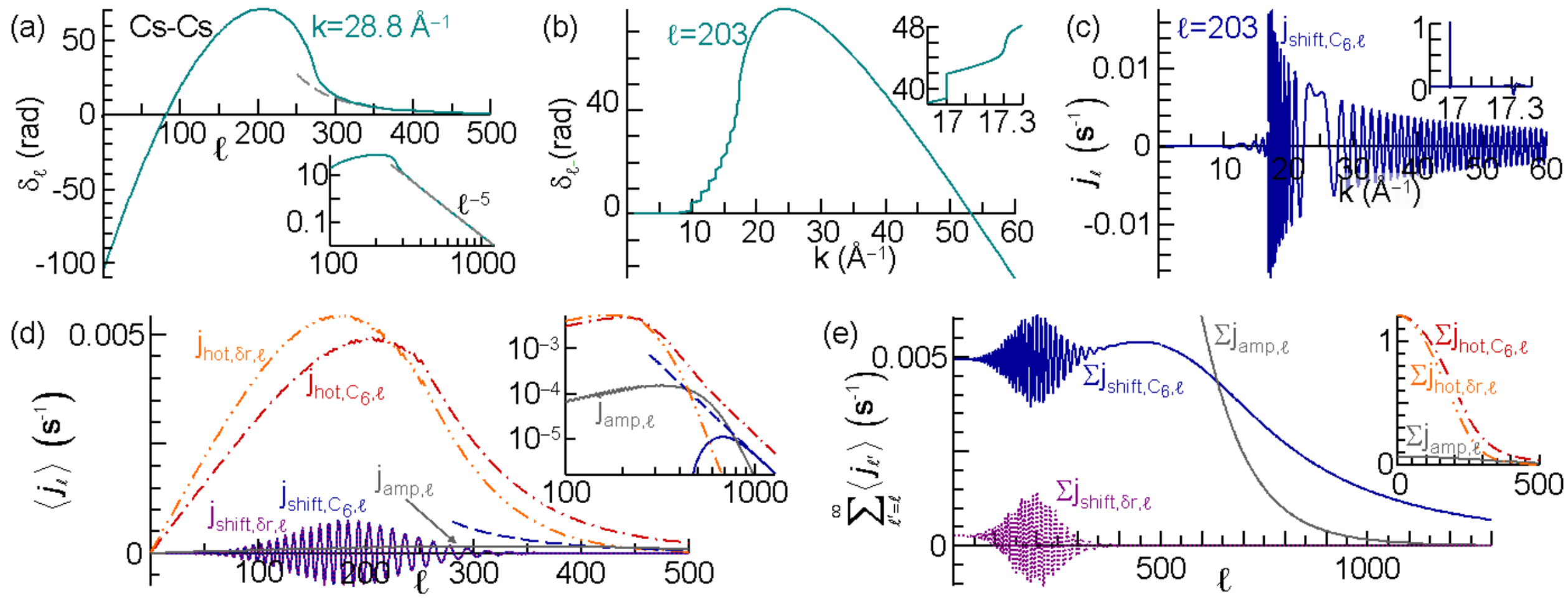

**Fig. 2.** (a) Partial wave scattering phase shifts for a model Cs-Cs scattering potential for the most probable speed at room temperature, $k_u$=28.8 Å$^{-1}$. The inset shows the asymptotic van der Waals $\ell^{-5}$ scaling. (b) Energy dependence of scattering phase shift for $\ell$ =203. At low k, there are Fano shape resonances, shown in the inset at 17.0 and 17.3 Å$^{-1}$, which produce rapid variations of the scattering phase. (c) Energy dependence of the $\ell$ =203 frequency shift of cold Cs clock atoms for a background Cs density n=10$^7$ cm$^{-3}$. The stationary phase near k=24 Å$^{-1}$ contributes a non-zero shift while the fast oscillations and the narrow Fano resonances in (c) average to 0. d) Thermally averaged frequency shifts for cold and hot Cs clock atoms, for long-range perturbations $\Delta C_6$ and short-range $\delta r$ (see text). The decrease in the Ramey fringe amplitude $j_{amp,C6,\ell}$ is also shown. In the inset, $j_{shift,C6,\ell}$ oscillates for $\ell$<480 and the shift for short range perturbations $j_{shift,\delta r,\ell}$ is negligible for large $\ell$. (e) Running sum from $\ell$ to ∞ of the thermal averages in (d). Long-range perturbations $\Delta C_6$ produce a cold atom shift at large $\ell$ whereas cold atoms are immune to short-range perturbations $\delta r$. Small $\ell$'s dominate the shifts for hot atoms and the loss of the cold atom Ramsey fringe amplitude.

$$
\begin{aligned}
j_{2,\mathrm{int}} &= -j_{shift}\sin(\phi) - j_{amp}\cos^2\left(\frac{\phi}{2}\right) \\
&= \frac{\pi n\hbar}{2\mu k}\sum_{\ell}(2\ell+1)\Big(\left[\sin(2\delta_{1\ell})-\sin(2\delta_{2\ell})\right]\sin(\phi) \\
&\quad -4\left[\sin^2(\delta_{1\ell})+\sin^2(\delta_{2\ell})\right]\cos^2\left(\frac{\phi}{2}\right)\Big)
\end{aligned} \quad . \qquad (1)
$$

Here, μ is the reduced mass, $\delta_{\gamma\ell}$ is the partial wave phase shift for clock state γ, n is the background gas density, and k is the relative wave-vector. The sin(ϕ) term $j_{shift}$ is the frequency shift. The $\cos^2(\phi/2)$ term $j_{amp}$ gives the decrease of the cold atom Ramsey fringe amplitude, as in the attenuation of a beam while passing through gas. In a room-temperature clock, including cell clocks [10], the sum of the scattered and interference currents of the clock atoms is, as we derive below:

$$
\begin{aligned}
j_{2,hot} = \frac{\pi n\hbar}{2\mu k}\sum_{\ell}(2\ell+1)\big[&\sin(2\delta_{1\ell}-2\delta_{2\ell})\sin(\phi) \\
&-2\sin^2(\delta_{1\ell}-\delta_{2\ell})\cos(\phi)\big]
\end{aligned} \quad . \qquad (2)
$$

We can see the dominant physics from (1) and (2). For cesium clock atoms, the phase shifts are nearly the same for long-range collisions so $\delta_{2\ell}=\delta_{1\ell}+\Delta\delta_\ell$, with $\Delta\delta_\ell$ small. Thus, when $\delta_{1,2\ell}>1$, the frequency shift $\sin(2\delta_{1\ell})-\sin(2\delta_{2\ell})\approx-2\Delta\delta_\ell\cos(2\delta_{2\ell})$ is on average zero, although these collisions do reduce the Ramsey fringe amplitude $1-j_{amp}$. However, in (2) the frequency shift is proportional to $\sin(2\delta_{1\ell}-2\delta_{2\ell})$ and does not average to zero since $\Delta\delta_\ell$ is not large for most $\ell$. Thus, for partial waves with large $\delta_{\gamma\ell}$, the frequency shift (1) is well suppressed for cold clock atoms that remain cold after the scattering, as compared to the shift (2) of room-temperature clock atoms. When $\delta_{\gamma\ell}$ is small, the contributions to each are the same. We next give simple analytic behaviors for the scattering and then explicitly illustrate these behaviors with model potentials.

Long-range $r^{-6}$ van der Waals interactions dominate the scattering for many systems, including ground-state Cs and neutral perturbers. A WKB approximation gives asymptotic scattering phase shifts for a $-C_6r^{-6}$ interaction, $\delta_\ell=3\pi\mu C_6k^4/16\hbar^2[\ell(\ell+1)]^{5/2}$. The phase shift grows quickly as $\ell$ decreases and k increases, corresponding to smaller impact parameters $b\approx\ell/k$ and stronger interactions.

The $C_6$'s of the two ground-state hyperfine components will be very similar, but not identical because the detunings with the excited states are slightly different. Expressing $C_6$ as a sum over oscillator strengths, we can bound the difference $\Delta C_6/C_6$ to be less than $h\nu[E_1^{-1}+(E_1+E_p)^{-1}]$, where $h\nu$ is the hyperfine splitting and $E_1$ and $E_p$ are the lowest resonant excitation energies of cesium and the perturber. For background Cs atoms, this gives $\Delta C_6/C_6$ less than 1/25000 and 1/34,000 for $H_2$. State-of-the-art calculations [18,19] can give more accurate $\Delta C_6$'s, but this illustrative and simple bound is useful. With it, we now derive general background-gas frequency shifts for cold atoms.

Using the WKB asymptotic scattering phase shifts, we sum (1) over all $\ell$ and average over the thermal speed distribution $\exp(-k^2/k_u^2)d\mathbf{k}^3$, with $\hbar^2k_u^2=2\mu^2k_BT/m_p$ for perturbers of mass $m_p$. This sum and thermal average are analytic to leading order:

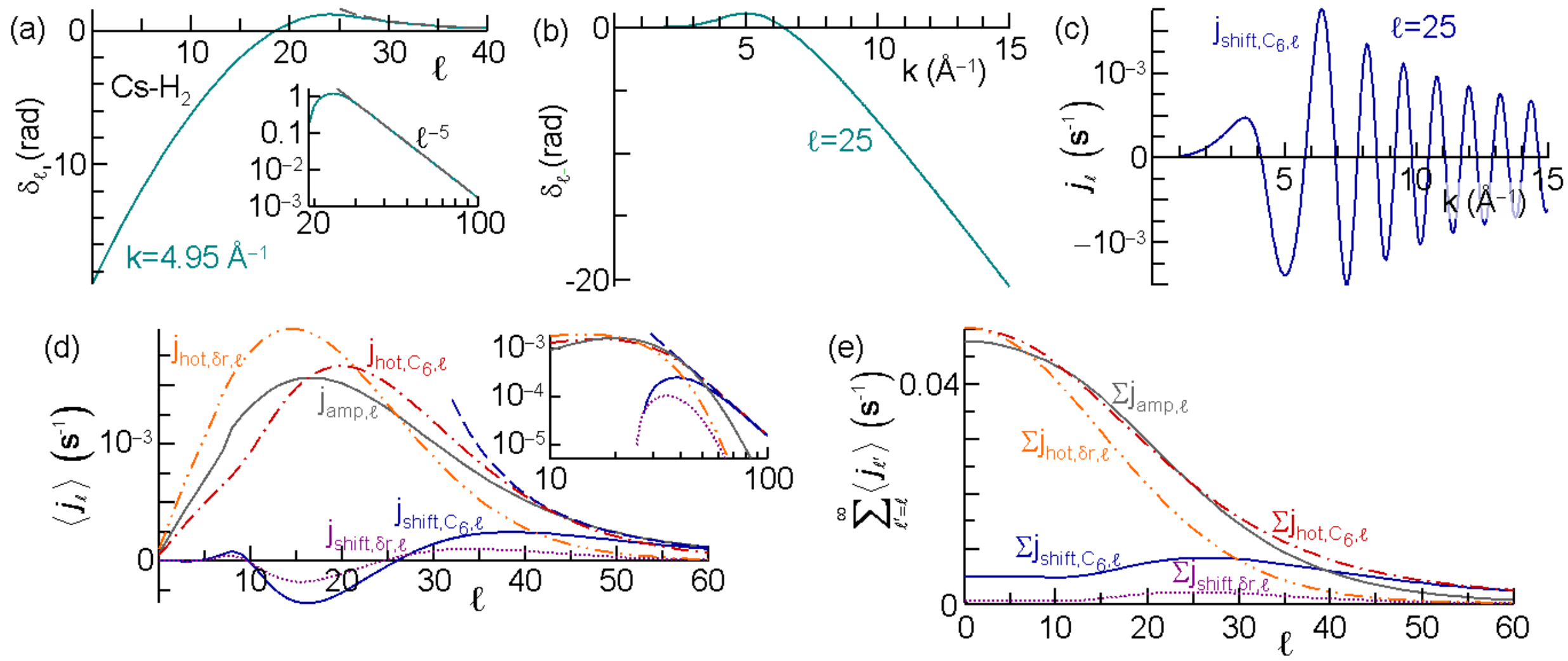

**Fig. 3.** Cs-$H_2$ model potential scattering as in Fig. 2. Here, $H_2$ is far less polarizable so the long-range van der Waals interaction produces smaller phase shifts in (a-c). The most probable speed of $H_2$ gives $k_u$=4.95 $Å^{-1}$. (d-e) Even though the van der Waals phase shifts are not large, the cold atom frequency shift is still suppressed for short-range perturbations δr (see text).

$$\sum_\ell \langle j_{2,\mathrm{int}}\rangle \approx -0.62n\left(\frac{2\mu k_B T}{m_p^2}\right)^{3/10}\frac{C_6^{2/5}}{\hbar^{2/5}}\left[\frac{\Delta C_6}{C_6}\sin(\phi)+13.8\cos^2\left(\frac{\phi}{2}\right)\right], \quad (3)$$

with numerical constants given to the indicated accuracy.

In (3) it is striking that the ratio of the frequency shift current to the attenuation of the Ramsey fringe amplitude is $\Delta C_6/13.8C_6$. Thus, clock frequency shifts from background gas collisions can be very generally bounded by measuring the associated decrease of the Ramsey fringe amplitude. If background Cs and $H_2$ eject fewer than ΔA=20% of the cold clock atoms during the $T_R$=0.5s Ramsey interrogation, the fractional frequency shift $-\Delta\nu/\nu=\Delta A/(13.8\pi\nu T_R)\Delta C_6/C_6$ must be less than $4\times10^{-17}$, which currently contributes insignificantly [1,2]. We note that this limit is the ratio of a frequency shift for collisions that have large impact parameters ($\delta_\ell$<1) to the attenuation from collisions with $\delta_\ell$>1 (small impact parameters), where $\sin^2(\delta_{\nu\ell})$ averages to ½. It is general, provided that the power law interaction is sufficiently strong to dominate other interactions for $\delta_\ell \lesssim 1$.

*Explicit calculations*: We now illustrate the behaviors of (1-3) with model potential surfaces for background $H_2$ and Cs vapor, the two most likely background gases in Cs fountain clocks. For Cs-Cs scattering we use $V(r)=-C_6/r^6+C_{12}/(r+3.2+\delta r)^{12}$ with $C_6$=6890 in atomic units (a.u.) [20] and $C_{12}$= $1.8535\times10^{11}$ a.u., to reasonably represent the $Cs_2$ triplet potential of [21]. Here we parameterize a short-range perturbation δr in V(r). Qualitatively, during short-range interactions, the perturber electrons tend to push the valance electron to overlap more with the nucleus, leading to a positive frequency shift for room-temperature clock atoms. For long-range attractive van der Waals interactions, the valence electron's overlap with the nucleus is reduced and therefore the hyperfine frequency decreases [10]. For $H_2$ we use a 6-12 model potential with $C_6$=170 a.u. [22] and $C_{12}=6.26\times10^7$ a.u., approximating a hard core estimate from a density functional calculation.

Fig. 2(a) shows the partial wave phase shifts for the most probable perturber speed $\hbar k_u/\mu$ for Cs-Cs scattering. The attractive long-range van-der Waals interaction gives the asymptotic phase shift above for large ℓ (gray dashed line) and $\delta_\ell$ exceeds 1 comfortably in the $Cs_2$ long-range van der Waals potential. At small ℓ the phase shift turns negative for the repulsive hard-core scattering. Versus k, the phase shift in Fig. 2(b) for ℓ=203 has a maximum near k=24 $Å^{-1}$ and a series of phase jumps through Fano-shape resonances at k=9 to 18 $Å^{-1}$ (inset). Fig. 2(c) shows the corresponding frequency shift from background gas collisions. At low k, the angular momentum barrier gives small scattering phase shifts and small frequency shifts. At high k, the frequency shift oscillates as the phase shift successively wraps through 2π. Near k=24 $Å^{-1}$ the phase is stationary, leading to a positive contribution to the thermally averaged frequency shift for ℓ=203. The narrow Fano resonances produce large positive and negative frequency shifts versus k [Fig. 2(c) inset], but are insignificant after thermal averaging. The frequency shift $j_{shift,C6,\ell}$ is linear in $\Delta C_6$ and scaled so that multiplying it by $-\Delta C_6/C_6$ gives the actual shift.

Fig. 2(d) shows the thermally averaged scattered probability currents for cold and hot-atom background gas collisions. As ℓ decreases, the maximum phase in Fig. 2(b) increases, and therefore the frequency shifts from the stationary region in Fig. 2(c) oscillates, producing the oscillation of $\langle j_{shift,C6,\ell}\rangle$ in Fig. 2(d). For large ℓ the shift for cold atoms asymptotically follows $\ell^{-5}$ (inset). The shift for hot-atoms $j_{hot,C6,\ell}$ does not oscillate with ℓ, increasing as ℓ decreases, until the phase shift difference becomes large here, while more realistically the interactions are no longer $r^{-6}$. The gray curve shows the attenuation of cold atom Ramsey fringe $j_{amp,C6,\ell}$, which too is large for moderate $\ell$ and then falls as $\ell^{-10}$. We also show the shifts for short range perturbations $j_{shift,\delta r,\ell}$ and $j_{hot,\delta r,\ell}$. These are linear in $\delta r$ and normalized so that the magnitude of the total thermally averaged shift for hot atoms is the same as for $j_{hot,C6,\ell}$. The shift for $\delta r$ is biased to smaller $\ell$ (small impact parameters) for hot atoms as compared to $j_{hot,C6,\ell}$. For cold atoms, the normalized sensitivity is essentially the same for moderate $\ell$, again showing the sensitivity to the stationary phase as in Fig. 2(c), but then goes to zero for large $\ell$, which do not probe short-range perturbations.

Fig. 2(e) shows the running sum from $\ell$ to $\infty$ of the thermal averages in Fig. 2(d). We see that the oscillations give no net contribution because the stationary phase in Fig. 2(b) sweeps smoothly versus $\ell$. Therefore, short range perturbations $\delta r$ give a negligible shift, especially considering that it is further suppressed because the long-range hot shift $\Sigma\langle j_{hot,C6,\ell}\rangle$ is much larger than $\Sigma\langle j_{hot,\delta r,\ell}\rangle$. Here we indeed see that $\Sigma\langle j_{shift,C6,\ell}\rangle$ gets all of its contributions from large $\ell$ and the ratio $\Sigma\langle j_{shift,C6,\ell}\rangle/\Sigma\langle j_{amp,C6,\ell}\rangle$ is 13.3, close to 13.8 in (3). Note that the shift for hot atoms from this model potential, essentially $\Delta C_6/C_6\Sigma\langle j_{hot,C6,\ell}\rangle$, overestimates the shift for cold atoms $\Delta C_6/C_6\Sigma\langle j_{shift,C6,\ell}\rangle$, by a factor of 200. From the sum over all $\ell$ in Fig. 2(e), a background Cs density $n=10^7$ $cm^{-3}$ gives $-\Delta\nu/\nu=\Delta C_6/\pi\nu C_6\Sigma\langle j_{shift,C6,\ell}\rangle <6.8\times10^{-18}$.

Fig. 3 shows results as in Fig. 2 for our Cs-$H_2$ model potential. Here, in comparison to Cs-Cs scattering, $C_6$ is small enough that the scattering phase shifts do not become large in the van-der-Waals regime of the potential surface and the range of contributing $\ell$'s is correspondingly smaller. The shift $j_{shift,\delta r,\ell}$ from a short range perturbation $\delta r$ in Fig. 3(d) is now slightly different than $j_{shift,C6,\ell}$, but again oscillates along with the stationary phase versus k. He and $H_2$ have the smallest polarizabilities and these are sufficiently small that the van der Waals phase shifts are not large. Thus, short-range and $\Delta C_6$ perturbations both contribute in the same range of $\ell$'s [Fig. 3(d) inset]. Despite the moderate van-der-Waals phase shifts, the ratio $\Sigma\langle j_{shift,C6,\ell}\rangle/\Sigma\langle j_{amp,C6,\ell}\rangle$ in Fig. 3(e) is 9.6, not so much less than 13.3 for a highly polarizable Cs gas, and gives $-\Delta\nu/\nu<4\times10^{-17}$ for $\Delta A<20\%$. Again the model shifts for hot atoms overestimates the shift for cold atoms, by a factor of 10 for $\Delta C_6$ and 94 for $\delta r$. The sum over all $\ell$ in Fig. 3(e) gives $-\Delta\nu/\nu=\Delta C_6/\pi\nu C_6\Sigma\langle j_{shift,C6,\ell}\rangle <5.1\times10^{-18}$ for $n=10^7$ $cm^{-3}$. It has the opposite sign and is an order of magnitude smaller than the measured room-temperature clock shift [10], which has partially cancelling short and long-range perturbations.

In ion clocks, the ions induce a dipole moment of the neutral background gas yielding a strong $r^{-4}$ interaction. The differential clock state perturbation occurs through the dipole-induced-dipole $r^{-6}$ interaction, giving a small $\Delta C_6$, which will have negligible consequences for the interference frequency shift (1), due to the large $r^{-4}$ phase shifts. For lattice clocks, $r^{-6}$ interactions dominate, but with a larger $\Delta C_6$, not significantly suppressed by the energy difference of the $^1S_0$ and $^3P_0$ clock states as compared to the energies of their strongest resonances. More polarizable clock atoms, including Sr and Yb, have naturally larger cross sections in (3).

*Scattering of Coherent Superpositions*: Before scattering, a $\pi/2$ pulse prepares clock atoms in a coherent superposition of internal states $|1\rangle$ and $|2\rangle$. After a background atom collides with a clock atom, with center-of-mass momentum $k\hat{z}$, its wavefunction is $|\Psi\rangle=2^{-1/2}[\exp(ikz)+f_1(\theta)\exp(ikr)/r]|1\rangle+2^{-1/2}[\exp(ikz)+f_2(\theta)\exp(ikr)/r]|2\rangle$, in a frame rotating at the clock frequency, where $f_\gamma(\theta)$ is the scattering amplitude of a spherically symmetric potential [23]. Applying the second Ramsey $\pi/2$ pulse with phase $\phi$, the excited state amplitude is $\langle 2|\Psi\rangle=\frac{1}{2}[\exp(ikz)+f_1(\theta)\exp(ikr)/r]+\frac{1}{2}\exp(-i\phi)[\exp(ikz)+f_2(\theta)\exp(ikr)/r]$.

The excited state probability current density is $\hbar/\mu\,\mathrm{Im}(\langle\Psi|2\rangle\nabla\langle 2|\Psi\rangle)$. In the forward direction the cross term between exp(ikz) and exp(ikr) as $r\rightarrow\infty$ yields the interference current $j_{2,int}$. Using $f_\gamma(\theta)$ and summing over distributed scatterers gives (1). The scattered current $j_{2,sc}(\theta)$ equals $\hbar k/2\mu(\mathrm{Im}[f_1(\theta)f_2^*(\theta)]\sin(\phi)+2\mathrm{Re}[f_1(\theta)f_2^*(\theta)]\cos^2(\phi/2)+2|f_1(\theta)-f_2(\theta)|^2)$. We get (2) by adding $j_{2,sc}$ and $j_{2,int}$ and integrating over all solid angles, using the orthogonality of $P_{\ell'}(x)P_\ell(x)$.

The scattered current in the forward direction $j_{2,sc}(0)d\Omega$ is non-zero, but small relative to $j_{2,int}$. In fountain clocks, velocity changes of $v_{max}\approx2$ cm/s eject atoms from the fountain so that they are not detected. This limits $\delta\theta$ to $m_{Cs}v_{max}/\hbar k\lesssim1$ mrad. For this $\delta\theta$, $P_\ell[\cos(\theta)]\approx1$ for $\ell\lesssim1500$ [5], giving $j_{2,sc}(\theta\approx0)d\Omega\approx\frac{1}{4}\delta\theta^2 j_{shift}\sin(\phi)\Sigma_{\gamma,\ell}(2\ell+1)\sin^2(\delta_{\gamma\ell})+\ldots$ [24]. Using the WKB phase shifts, we get a thermally averaged $\langle j_{2,sc}(0)d\Omega\rangle=0.31\langle j_{shift}\rangle(m_p\mu C_6^2/\hbar^{14}k_u^2)^{1/5}m_{Cs}^2v_{max}^2$. For Cs or $H_2$ background gases and $v_{max}\approx2$ cm/s, the shift from scattered atoms is less than 1% of that from $j_{shift}$.

In summary, the frequency shifts of laser-cooled atomic clocks due to room-temperature background gases is an order of magnitude or more smaller than the shifts for room-temperature clocks. Thus, the significant uncertainty from background gas collisions in the most accurate cesium fountains clocks can become negligible. For cold clock atoms, that are post-selected to remain cold after scattering, the interference in the forward direction between the scattered and unscattered atomic waves dominates the frequency shift. Analytic and model potential calculations show that weak long-range, and not strong short-range interactions produce frequency shifts. General considerations show that the decrease of the Ramsey fringe amplitude can bound clock frequency errors below the current $10^{-16}$ level. Essentially all background gases, including $H_2$ and perhaps He, have sufficient polarizability to exhibit these behaviors.

We acknowledge helpful conversations with M. Cole, I. Iordanov, J. Sofo, and K. Szymaniec, and financial support from NASA, NSF, and Penn State.